\documentclass[british]{eptcs}
\usepackage[T1]{fontenc}
\usepackage[latin9]{inputenc}
\usepackage{color}
\usepackage{refstyle}
\usepackage{url}
\usepackage{algorithm2e}
\usepackage{amstext}
\usepackage{graphicx}
\usepackage{verbatim}
\usepackage{wrapfig}
\makeatletter

\newcommand{\iflong}[1]{}

\AtBeginDocument{\providecommand\tabref[1]{\ref{tab:#1}}}
\providecommand{\tabularnewline}{\\}

\RS@ifundefined{subsecref}
  {\newref{subsec}{name = \RSsectxt}}
  {}
\RS@ifundefined{thmref}
  {\def\RSthmtxt{theorem~}\newref{thm}{name = \RSthmtxt}}
  {}
\RS@ifundefined{lemref}
  {\def\RSlemtxt{lemma~}\newref{lem}{name = \RSlemtxt}}
  {}

\usepackage[all]{xy}

\makeatother

\usepackage{babel}
\usepackage{listings}
\begin{document}

\global\long\def\X{X}
\global\long\def\F{F}
\global\long\def\G{G}


\title{A Parallel Linear Temporal Logic Tableau}

\author{John C. M\textsuperscript{c}Cabe-Dansted, Mark Reynolds  \institute{School of Computer Science and Software Engineering\\
The University of Western Australia\\
Australia\\
\email{john.mccabe-dansted@uwa.edu.au}, \email{mark.reynolds@uwa.edu.au}} 
}

\def\authorrunning{John C. M\textsuperscript{c}Cabe-Dansted \&
Mark Reynolds}
\def\titlerunning{A Parallel Linear Temporal Logic Tableau}
	\maketitle
\begin{abstract}
For many applications,
we are unable to take full
advantage of the potential massive parallelisation 
offered by supercomputers or cloud computing
because it is too hard to work out how to divide up the
computation task between
processors in such a way to minimise the
need for communication.
However, a recently developed branch-independent 
tableaux for the common LTL temporal logic
should intuitively be easy to parallelise as each branch
can be developed independently. Here we describe a simple technique for partitioning such a
tableau such that each partition can be processed independently
without need for interprocess communication. We investigate the extent
to which this technique improves the
performance of the LTL tableau
on standard benchmarks and random formulas.

\end{abstract}

\section{Introduction}

Parallelisation is important to exploit modern multi-core computers. There has been considerable interest in the potential of tableaux to be parallelised (see e.g. \cite{DBLP:conf/tableaux/Schwendimann98}, \cite{FLL10}, \cite{DBLP:conf/ijcai/BertelloGMR16}).
For this reason we propose and investigate a simple technique for parallelising tableaux with independent branches.

An embarrassingly easy problem is one that can be easily divided into
multiple tasks, particularly when there is no need for communication
between those tasks. The one-pass tableau of \cite{DBLP:conf/ijcai/BertelloGMR16}
should be an ``embarrassingly parallel problem'' as each branch
is independent. In principle, we can assign each branch to a different
thread, and report ``satisfiable'' if any of the tasks report that
the formula is satisfiable.

Nevertheless it is not axiomatic that improving performance via parallelization
is easy. In practice, the tableau may have many branches. Blindly
assigning a thread to every branch may flood a local machine. On cloud
services firing up as many CPUs as there are branches could be embarrassingly
expensive. In any case, moving problems to new CPUs has overhead,
and so assigning a branch to a new thread may not even improve performance.

A further problem when parallelising tableaux is that multi-threaded
programming can be hard. Since order of execution of parallel task
may not be deterministic, it can lead to ``Heisenbugs'' that are
hard to reproduce because they only occur occasionally even given
fixed input. Eliminating these bugs can be time consuming, and even
if all such bugs have been eliminated, it can be hard to trust the
correctness of proofs that depend on the correctness of the implementation.
Given a limited amount on time to implement and debug a tableau, it
may be wise to add more optimisations rather than implement complex
support for multi-threaded computation. 
The contributions of this paper are: 
\begin{itemize}
\item to introduce a novel, fast and easy way to manage parallel implementations
on suitable tableau reasoners which can make maximum use of the power
of multiple processors; 
\item to demonstrate the speed-up in practice using standard and very accessible
cloud based high performance and multi-core facilities; 
\item  to show that we can 
 predict 
the performance of the  technique on  unsatisfiable formulas;
\item to give an indication of what sorts of problems benefit most from
such an approach; and 
\item to suggest what role such parallel tableau system may play in combination
with a host of very different rival reasoning systems. 
\end{itemize}
In the next section we will quickly review some relevant details of
the recently introduced LTL tableaux, which has the one-pass and branch-independence
aspects that required for our parallelisation technique. In Section~\ref{sec:partitioner}, we introduce
the parallel algorithm which manages the partitioning of the tableau
search into separate jobs for the given number of parallel processors.
In Section~\ref{sec:Benchmarks}, we present some benchmarks of satisfiable
and unsatisfiable formulas. In Section \ref{sec:Analysis}, we consider the
shape of tableaux and how this affects when parallelisation is effective. In Section \ref{sec:compare}, we consider what our parallel implementation adds to the portfolio of reasoners with different strengths.
Finally, we present a brief conclusion mentioning future work. Some additional benchmarks are available in the expanded version \cite{parallel-long}.

 \newcommand{\webpage}{\url{http://staffhome.ecm.uwa.edu.au/~00054620/research/Online/ltlsattab.html}}
 \newcommand{\techreport}{DBLP:journals/corr/Reynolds16}
 \newcommand{\ijcai}{DBLP:conf/ijcai/BertelloGMR16}

\section{\label{sec:ltl}A new one-pass embarrassingly parallel LTL tableau}

In this section we give a brief introduction to the 
branch-independent LTL tableaux system
recently introduced
in \cite{DBLP:journals/corr/Reynolds16a}
and evaluated experimentally against other
state-of-the-art reasoners in
\cite{DBLP:conf/ijcai/BertelloGMR16}.

\begin{wrapfigure}{R}{0.5\textwidth}
	\begin{center}
		\includegraphics[width=8cm]{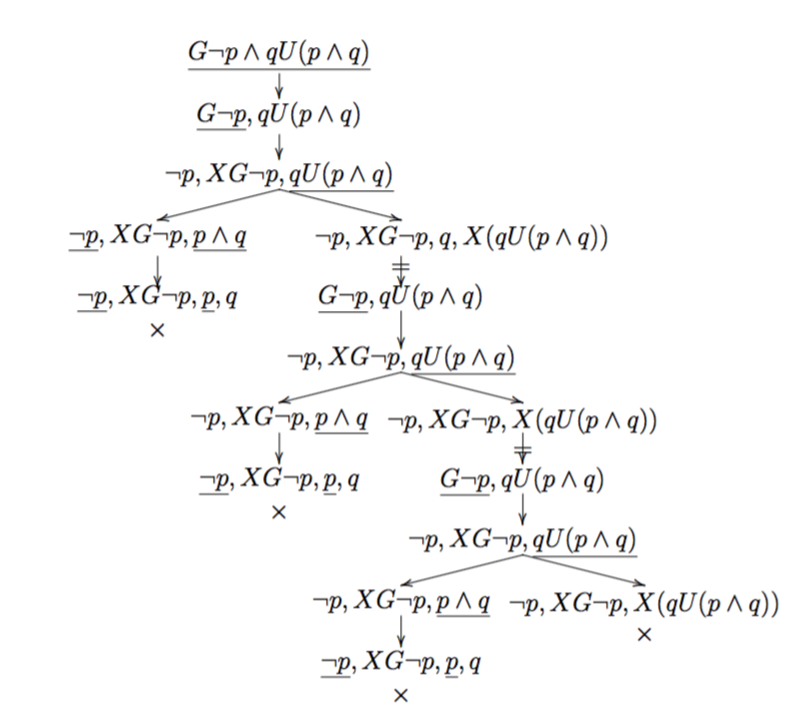}
	\end{center}
	\caption{Tableau Example using Prune Rule to close rightmost branch}
	\label{fig:feg}
\end{wrapfigure}

LTL \cite{Pnu77}, or propositional linear time temporal logic,
is a long established formalism for 
tackling a range of industrially important
reasoning tasks from the correctness
of hardware systems, 
through AI planning to
specifying work flow.
LTL satisfiability checking,
seeing if an LTL formula has a model or not,
has recently being receiving renewed attention \cite{DBLP:journals/corr/LiP0VH14}
in ``sanity checking'' of specifications:
there is no point trying to implement
an inconsistent specification.

An important property of LTL and the leviathan tableau is that they only consider one branch. This is all that our parallelisation technique depends upon. It does not otherwise interact the semantics of LTL; however, for completeness we now introduce the syntax of LTL.
LTL adds temporal operators such as tomorrow $X$, eventually $F$, always $G$ and until $U$
on top of classical propositional logic
and evaluates truth of formulas along paths of states,
where each proposition is true or false at the individual states.
 We define the (well-formed) formulas of LTL recursively: 
any atomic proposition is a formula, if $\alpha$ and $\beta$ are formulas then $\neg \alpha$, $\alpha \wedge \beta$, $\alpha \vee \beta$,  $X\alpha$, $F\alpha$, $G\alpha$ and $\alpha U \beta$ are all formulas of LTL.
See \cite{DBLP:journals/corr/Reynolds16a} for details of the semantics.
We say that a formula is satisfiable iff
there is a sequence of states where the formula
holds at the initial state when evaluated
along the whole path.

There are quite a range of rival reasoning techniques and tools which 
can be brought to bear on the
LTL satisfiability problem.
They include
tableaux pltl \cite{DBLP:conf/tableaux/Schwendimann98},
resolution TRP++ \cite{hustadt2003trp++},
resolution LS4 \cite{SuW2012},
symbolic model verification NuSMV \cite{Cimatti2002}, 
and automata Aalta \cite{LiY2014}.
Surprisingly (or is it unsurprisingly?), there are no clear overall winners.
Therefore, ``portfolio'' reasoners such as \cite{DBLP:journals/corr/LiP0YVH13}
have a part to play:
trying a variety of tools in parallel
on the same input formula.

As mentioned in the Introduction,
apart from the portfolio reasoners
and a very preliminary 
account of an idea in
\cite{parallel06},
parallel computing techniques
have not been
applied to LTL. \iflong{

} Recently a new tableau rule was introduced 
that allows a traditional tree-shaped tableau \cite{DBLP:journals/corr/Reynolds16a}.
It is a one-pass tree-shaped one like Schwendimann
\cite{DBLP:conf/tableaux/Schwendimann98}
but unlike Schwendimann's, there is no need for communication
between branches.

The new rule, called the Prune rule, can be added to fairly standard tableau construction
rules.
It is some sort of negative counterpart of the looping rule
which allows a branch to be closed successfully
if a label is repeated along the branch
and cures to all eventualities (required by the label)
are witnessed in between the two appearances
of the same label.
The new rule allows branches to be failed, i.e.
closed unsuccessfully,
just because they have become repetitious
without making progress
in witnessing cures
to eventualities.
The Prune rule
can be applied
when three occurrences of the same label
appear down one branch
with no new eventualities
being cured between the
second and third occurrence.

By using the new Prune rule in amongst
quite traditional tree-shaped classical and modal logic
tableau construction rules,
we can provide a sound and complete
tableau system for LTL.
The completeness proof is quite intricate.
Figure~\ref{fig:feg}
presents
a small example tableau.
The reader will need
to see 
\cite{DBLP:journals/corr/Reynolds16a}
for full details of the
rules,
diagram notation and
proofs.

A demonstration Java implementation
of the tableau system
is available at
\webpage{}
which allows users to understand the tableau building process
in a step by step way
but it is not
designed as a fast implementation.
A quite straightforward but fast and efficient C++ implementation of the new tableau
written by
Matteo Bertello
of Udine University
is available from 
\url{https://github.com/Corralx/leviathan}.
This is evaluated in \cite{DBLP:conf/ijcai/BertelloGMR16} against 
a wide range of state-of-the-art reasoners
using
benchmarks from \cite{VSchuppanLDarmawan-ATVA-2011}.

The Prune-based tableau is fast, 
and it is also intuitive and simple to use manually and implement.
It has some other advantages including
extensibility: for example
a clever but simple additional rule allows 
the tableau to also handle past-time operators
\cite{pltltab}.
The important aspects of this new pruning-based LTL tableau approach
for us are its one-pass nature and 
the fact that branches can be built and evaluated independently
of each other.
As noted in \cite{DBLP:journals/corr/Reynolds16a}
this suggests that the tableau search in the new tableau
is an ``embarrassingly parallel'' task
\cite{Fos95}.
Let us see if we can make use of this potential.

\section{\label{sec:partitioner}Parallelisation Technique}

\begin{wrapfigure}{R}{0.57\textwidth}
\begin{center}
\includegraphics[scale=0.70]{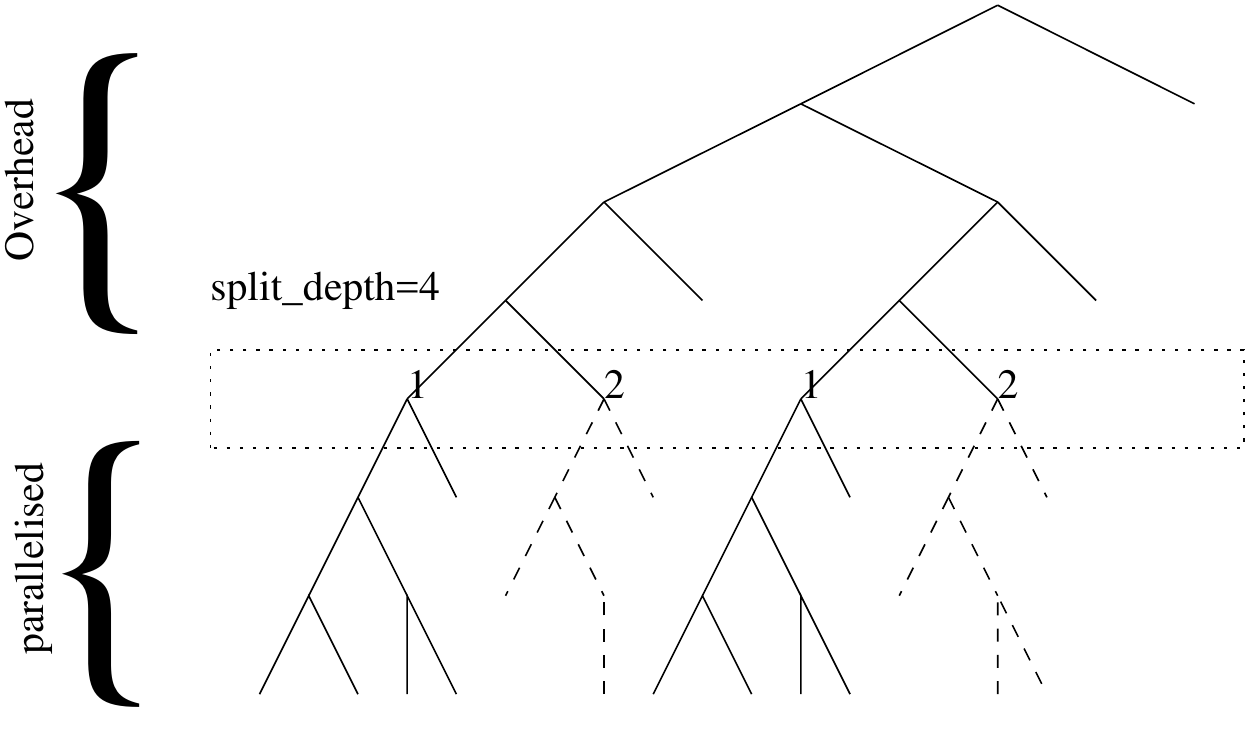}
\caption{\label{fig:jobtree}Portion of an example tableau constructed by job 1 out of 2. The portion of the tableau above \texttt{split\_depth} is constructed by all jobs. We have labelled vertices at depth \texttt{split\_depth} with a job number. At \texttt{split\_depth} job $j$ immediately rejects/fails the current branch if it is not labelled with $j$.}
\end{center}
\end{wrapfigure}

\begin{algorithm}
	\hrule{}
	
\begin{lstlisting}
int to_job (int i) {
	i=i-1;
	int m = (i %number_of_jobs);
	int r = (i /number_of_jobs);
	if (r > 0) {
		srand(r); /*set random seed*/
		return 1+((rand() + m) %number_of_jobs);
	} else {
		return 1+m;
	}
}
\end{lstlisting}
	
	\hrule{}\caption{\label{alg:to_job}{\tt to\_job}: Algorithm to assign the $i^{\text{th}}$ branch of given depth to a job. This algorithm can be informally described as ``Where $n$ is the number of jobs, break the jobs into blocks of size $n$. Assign each block of $n$ jobs in a round-robin fashion, but for each block start assigning jobs at some pseudo-randomly chosen job''.}
\end{algorithm}

\begin{algorithm}
\hrule{}

\begin{lstlisting}
int last_depth=0;
int width=0;

bool decline() {
  assert(get_depth() <= split_depth || job_no > 0);
  if (get_depth() > last_depth && get_depth() == width) {
    width++;
  }
  if (get_depth()==split_depth && last_depth<split_depth) {
    if (to_job(width)!=(job_no)) {
       while (get_depth() >= split_depth) 
             _rollback_to_latest_choice(); //fail branch
       last_depth=get_depth();
       return true;
    }
  }
  last_depth=get_depth();
  return false;
}
\end{lstlisting}
\hrule{}
\caption{\label{alg:decline}The Decline rule.  This Algorithm fails/rolls-back the current branch iff we are at {\tt split\_depth}, {\emph{and}} the present branch is not assigned to the current {\tt job\_no}.}

\end{algorithm}

\begin{algorithm}
\hrule
\begin{lstlisting}[language=bash]
SPLIT_DEPTH=1        #Sample Input
FORMULA="Xp&~Xp|XXq" #Sample Input

nCPU=`nproc`
seq 1 $nCPU |
    parallel --halt 2 JOB_NO={}/$nCPU@$SPLIT_DEPTH\\
    bin/checker -l "'$FORMULA'"
case $? in
0) echo VOTE: Unsatisfiable ;;
5) echo VOTE: Satisfiable ;;
*) echo Unknown Result
esac
\end{lstlisting}
\hrule{}
\caption{\label{alg:bash}Shell script wrapper to run jobs on each CPU of a single machine, {\tt bin/checker} is the {\tt leviathan} tableau, {\tt parallel -{}-halt 2} runs jobs in parallel until a non-zero exit code is returned, the {\tt JOB\_NO} environment variable has the format job-number/number-of-jobs@{\tt split\_depth}, {\tt \$?} is the exit code of the last command. This reports satisfiable as soon as any of the jobs reports satisfiable, or unsatisfiable once all jobs report unsatisfiable.
	}
\end{algorithm}

We parallise the tableau using another rule we call the Decline rule. 
The idea is that at some fixed ``{\tt split\_depth}'', we divide the branches between multiple independant ``jobs'' that can be run in parallel. The decline rule fails the present branch if it has been assigned to another job. 
As soon as \emph{any} job halts and reports that the formula is \emph{satisfiable}, the entire parallel algorithm halts and reports that the formula is satisfiable. If \emph{all} jobs have halted and reported
that the formula is \emph{unsatisfiable}, the parallel algorithm halts and reports to the user that the formula is unsatisfiable (if only some jobs report that the formula is unsatisfiable, 
this may just be because none of the model(s) of the  formula are reachable from vertices assigned to those jobs). 

The division of a tableau into jobs is illustrated in Figure \ref{fig:jobtree}. Parts of the tree above the {\tt split\_depth} are overhead that need to be constructed by all jobs, whereas parts of the tree below {\tt split\_depth} are parallelised. In Figure \ref{fig:jobtree} the jobs have been assigned in a round-robin fashion, and this has resulted in job 1 picking a left branch each time. We also see that this has resulted in more work being assigned to job 1. Although this figure is for illustrative purposes, we have found that using a simple round-robin assignment can lead to an imbalance in the amount of work assigned to each job. For this reason we use Algorithm \ref{alg:to_job} to assign branches to jobs. 

We have not shown the formulas on each branch of Figure \ref{fig:jobtree}. This is because this parallisation technique does not take as input formulas, it only works on the shape of the graph. The only property this parallelisation technique requires of the tableau is that the branches are independent.

\begin{table}
		\resizebox{\columnwidth}{!}{
			\begin{tabular}{|l|r@{.}l|r@{.}l|r@{.}l|r@{.}l|l|r@{.}l|r@{.}l|r@{.}l|}
				\cline{2-9}\cline{11-16}
				\multicolumn{1}{c}{}&\multicolumn{8}{|c|}{Time Taken}&&\multicolumn{6}{|c|}{speedup}\\
				\hline
				Name&\multicolumn{2}{|c|}{$\times 1$}&\multicolumn{2}{|c|}{$\times 2$}&\multicolumn{2}{|c|}{$\times 8$}&\multicolumn{2}{|c|}{$\times 88$}&formula&\multicolumn{2}{|c|}{$\times 2$}&\multicolumn{2}{|c|}{$\times 8$}&\multicolumn{2}{|c|}{$\times 88$}\\
				\hline		
				
				U3\_0002 & 27&16 & 13&49 & 3&93 & 0&59 & trp/N5x/32-5-0-32-3-0-200001 & 2&01 & 6&91 & 46&03\\
				U3\_0003 & 51&51 & 23&51 & 7&52 & 1&01 & trp/N5x/30-5-0-30-3-0-200003 & 2&19 & 6&85 & 51&00\\
				U3\_0004 & 52&39 & 25&50 & 8&54 & 1&06 & trp/N5x/40-5-0-40-3-0-200000 & 2&05 & 6&13 & 49&42\\
				U3\_0005 & 56&51 & 27&74 & 11&14 & 1&26 & trp/N5x/32-5-0-32-3-0-200002 & 2&04 & 5&07 & 44&85\\
				U3\_0006 & 74&16 & 36&51 & 12&15 & 1&51 & trp/N5x/35-5-0-35-3-0-200006 & 2&03 & 6&10 & 49&11\\
				U3\_0007 & 82&79 & 40&70 & 11&61 & 2&05 & trp/N5x/32-5-0-32-3-0-200009 & 2&03 & 7&13 & 40&39\\
				U3\_0008 & 118&41 & 59&93 & 19&55 & 3&66 & trp/N5x/30-5-0-30-3-0-200000 & 1&98 & 6&06 & 32&35\\
				S3\_0006 & 13&67 & 6&09 & 0&03 & 0&02 & rozier/.../P0.7N1L40\_6 & 2&24 & 455&67 & 683&50\\
				S3\_0007 & 16&64 & 16&66 & 16&46 & 0&63 & trp/N12y/36-12-0-36-3-0-200003 & 1&00 & 1&01 & 26&41\\
				S3\_0011 & 17&53 & 17&24 & 17&18 & 1&13 & trp/N12y/36-12-0-36-3-0-200004 & 1&02 & 1&02 & 15&51\\
				S3\_0012 & 16&61 & 7&96 & 1&01 & 0&02 & rozier/.../P0.5N1L30\_5 & 2&09 & 16&45 & 830&50\\
				S3\_0024 & 117&23 & 48&52 & 46&12 & 45&89 & rozier/.../counterCarryLinear12 & 2&42 & 2&54 & 2&55\\
				S3\_0026 & 35&42 & 34&75 & 34&74 & 1&38 & trp/N12y/36-12-0-36-3-0-200006 & 1&02 & 1&02 & 25&67\\
				S3\_0029 & 48&33 & 48&98 & 48&51 & 0&06 & trp/N12y/58-12-0-58-3-0-200004 & 0&99 & 1&00 & 805&50\\
				S3\_0032 & 43&92 & 21&87 & 3&45 & 0&03 & rozier/.../P0.333...N1L60\_7 & 2&01 & 12&73 & 1464&00\\
				S3\_0033 & 128&02 & 48&54 & 44&42 & 42&58 & rozier/.../counterCarry12 & 2&64 & 2&88 & 3&01\\
				S3\_0034 & 56&69 & 55&40 & 56&92 & 0&07 & trp/N12y/46-12-0-46-3-0-200003 & 1&02 & 1&00 & 809&86\\
				S3\_0038 & 75&14 & 73&54 & 73&44 & 37&59 & trp/N12y/36-12-0-36-3-0-200005 & 1&02 & 1&02 & 2&00\\
				S3\_0042 & 94&67 & 23&75 & 0&08 & 0&02 & rozier/.../P0.5N1L80\_4 & 3&99 & 1183&38 & 4733&50\\
				S3\_0043 & 115&33 & 114&48 & 114&75 & 0&04 & trp/N12y/66-12-0-66-3-0-200002 & 1&01 & 1&01 & 2883&25\\
				S3\_0044 & 76&23 & 33&41 & 0&01 & 0&02 & rozier/.../P0.95N1L80\_1 & 2&28 & 7623&00 & 3811&50\\
				\hline
			\end{tabular}
		}

		\caption{\label{tab:Performance}Speed of U3/S3 formulas with {\tt split\_depth} of 18 and divided into 1, 2, 8, or 88 parallel jobs. 
		}
	\end{table}
	

 
 
We see that the algorithms are fairly simple. For example, Algorithm \ref{alg:bash} is roughly as lengthy as Algorithms \ref{alg:to_job} and \ref{alg:decline}, despite Algorithm \ref{alg:bash} being just a wrapper around our modified {\tt leviathan} and GNU {\tt parallel} \cite{Tange2011a}. Indeed Algorithm \ref{alg:bash} is a greatly simplified version of the actual script {\tt parallel.sh} that we use to divide jobs between instances, whereas Algorithms \ref{alg:to_job} and \ref{alg:decline} are very similar to the actual C++ code. Additionally we see that they are not tightly coupled to the implementation of {\tt leviathan}. Only 51 new lines of code were added to the C++ code in total, including
instrumentation. This increased the size of the file {\tt solver.cpp} which implements the tableau algorithm to
608 lines of code in total. The full implementation is available at
\url{https://github.com/gmatht/leviathan}, along with links to raw benchmark data.

Despite its simplicity, this approach has a number of advantages. 
Each job is a simple single threaded task, avoiding the potential for ``Heisenbugs'' that can arise from multi-threading and other forms of non-deterministic parallelism.
Since each job does not need to communicate with other jobs, we can
easily run jobs on different cloud instances. For example, the largest
single instance available on Amazon is the 128 vCPU {\tt m4.16xlarge} with 64 cores. 
In this paper we will discuss distributing tasks over more cores than available on a single EC2 instance.
By measuring the amount of time each vertex takes to process, we can
estimate the amount of time a formula would take to process if we
had used a different number of cores.
We can reproduce satisfiability results easily, including time required,
on a single core by re-running the job that determined the formula
was satisfiable. (Likewise, to demonstrate that the algorithm is slow
for a known unsatisfiable formula, we only need to show that
one job is slow.)
\iflong{
We assign vertices to jobs using Algorithm \ref{alg:to_job}. We prevent
the tableau from going past vertices not assigned to the current process
using \ref{alg:partition_tableau}; assigning job\_no to 0 prevents
a process from going past any of the vertices at {\tt split\_depth}. This is
primarily useful for measuring the overhead of the duplicated work
each process has to do.} 

\section{\label{sec:Benchmarks}Benchmarks}
\begin{table}
	\begin{tabular}{|l|l|r@{.}l|r@{.}l|r@{.}l|r@{.}l||l|l|r@{.}l|r@{.}l|r@{.}l|r@{.}l|}
		\hline
		\multicolumn{10}{|c||}{{\tt split\_depth}=18} &
		\multicolumn{10}{|c|}{{\tt split\_depth}=20} \\
		\hline
		jobs&set&\multicolumn{2}{|c|}{mean}&\multicolumn{2}{|c|}{median}&\multicolumn{2}{|c|}{min}&\multicolumn{2}{|c||}{max}&
		jobs&set&\multicolumn{2}{|c|}{mean}&\multicolumn{2}{|c|}{median}&\multicolumn{2}{|c|}{min}&\multicolumn{2}{|c|}{max}\\
		\hline
		2 & U2 & 1&97 & 1&98 & 1&92 & 2&00 & 2 & U2 & 1&83 & 1&84 & 1&70 & 1&91\\
		8 & U2 & 4&71 & 4&60 & 4&42 & 5&24 & 8 & U2 & 5&16 & 4&76 & 4&00 & 7&10\\
		32 & U2 & 15&14 & 15&34 & 11&50 & 17&91 & 32 & U2 & 12&47 & 13&52 & 7&93 & 16&41\\
		88 & U2 & 22&94 & 21&39 & 13&60 & 34&03 & 88 & U2 & 18&14 & 19&84 & 8&52 & 25&05\\
		\hline
		2 & U3 & 2&04 & 2&03 & 1&98 & 2&19 & 2 & U3 & 1&99 & 2&00 & 1&85 & 2&05\\
		8 & U3 & 6&19 & 6&12 & 5&07 & 7&13 & 8 & U3 & 6&80 & 7&24 & 5&17 & 8&42\\
		32 & U3 & 20&24 & 20&12 & 17&25 & 23&31 & 32 & U3 & 19&93 & 20&29 & 18&20 & 21&55\\
		88 & U3 & 44&17 & 45&44 & 32&35 & 51&00 & 88 & U3 & 44&92 & 45&27 & 38&37 & 48&06\\
		\hline
		2 & S2 & 1&08 & 1&01 & 0&98 & 2&07 & 2 & S2 & 1&02 & 0&98 & 0&84 & 1&95\\
		8 & S2 & 1&50 & 1&01 & 0&81 & 12&44 & 8 & S2 & 1&49 & 1&01 & 0&82 & 10&25\\
		32 & S2 & 4&65 & 1&01 & 0&90 & 90&50 & 32 & S2 & 3&90 & 1&01 & 0&75 & 67&88\\
		88 & S2 & 14&86 & 1&01 & 0&91 & 543&00 & 88 & S2 & 8&02 & 1&01 & 0&59 & 181&00\\
		\hline
		2 & S3 & 1&25 & 1&02 & 0&98 & 3&99 & 2 & S3 & 1&25 & 1&01 & 0&99 & 3&60\\
		8 & S3 & 207&45 & 1&02 & 0&96 & 7623&00 & 8 & S3 & 202&07 & 1&02 & 0&91 & 7623&00\\
		32 & S3 & 449&40 & 1&02 & 0&69 & 9467&00 & 32 & S3 & 449&78 & 1&02 & 0&96 & 9467&00\\
		88 & S3 & 358&44 & 1&04 & 0&89 & 4733&50 & 88 & S3 & 267&30 & 1&02 & 0&93 & 4733&50\\
		\hline
	\end{tabular}

	\caption{\label{tab:Speedup-factor}Speedup factor parallel algorithm over serial algorithm}
\end{table}


We consider the widely used \cite{DBLP:conf/ijcai/BertelloGMR16, DBLP:journals/corr/LiP0YVH13} LTL benchmark sets: {\tt acacia}, {\tt   alaska}, {\tt   anzu}, {\tt   forobots}, {\tt   rozier}, {\tt   schuppan}, and {\tt trp}. We identify the formulas by filename, and include them in our fork of {\tt leviathan} at  \url{https://github.com/gmatht/leviathan/tree/master/tests}.

We conducted a preliminary
study on these benchmarks, running them locally on an i7-4790 CPU
@ 3.60GHz running an Ubuntu 16.04 VirtualBox under Windows 7 using default settings (that is, without the {\tt -{}-release} switch, which increased performance roughly ten fold). Based on this study we divided
these formulas into {\tt U}nsatifiable and {\tt S}atisfiable, and 4 categories of difficultly, from (0): problems
which took between $10^{-1}$ to $10^{0}$ seconds to solve through
to (3): problems which took between $10^{2}$ to $10^{3}$ seconds
to solve. We exclude problems easier than (0) as being too easy to
consider parallelising and problems harder the (3) as being too hard
to benchmark effectively. This gave us 8 benchmark sets, U0--3 and S0--3. For
the cloud benchmarks we used Amazon {\tt c4} instances
 with Intel Xeon E5-2666 v3 CPUs.

In this paper we will use the term {\em vertices} to refer to the nodes of the tableau, i.e. the end points of a branch. We will use Amazon's term {\em instance} to refer to a computational node, i.e. a shared memory machine with one or more CPUs.

LTL formulas are commonly run on StarExec (\url{https://www.starexec.org/}), but to easily scale to
large numbers of CPUs we use Amazon Elastic Compute Cloud (EC2, \url{https://aws.amazon.com/ec2/}). On
EC2 we use the modern Compute Optimized ({\tt c4}) instances with Intel
Xeon E5-2666 v3 (Haswell) CPUs. 
\iflong{
	The largest cluster we consider is
of four {\tt c4.x8large} instances, each providing 36 vCPUs for a total
of 144 vCPUs. Our algorithm requires as input a fixed depth. We will
use 18 as the default.
} 
\iflong{Despite being ``compute optimised'' the {\tt c4} instances are less powerful
than the desktop i7 CPU.}
The {\tt c4} instances use hyper-threading
to provide twice as many vCPUs as CPUs, which slows each job by about 65\%. Overall each
vCPU is roughly half as powerful as a single thread of the desktop
i7. 
In this paper, we only run one job per two vCPUs on the instance. 
Although hyperthreading can provide more performance per physical CPU, this complicates the benchmarks. Benchmarks exploiting all the vCPUs are considered in the expanded version \cite{parallel-long}.

\subsection{\label{sec:sat}Satisfiable formulas}

\begin{figure}
	\begin{center}
		\scalebox{0.89}{\includegraphics{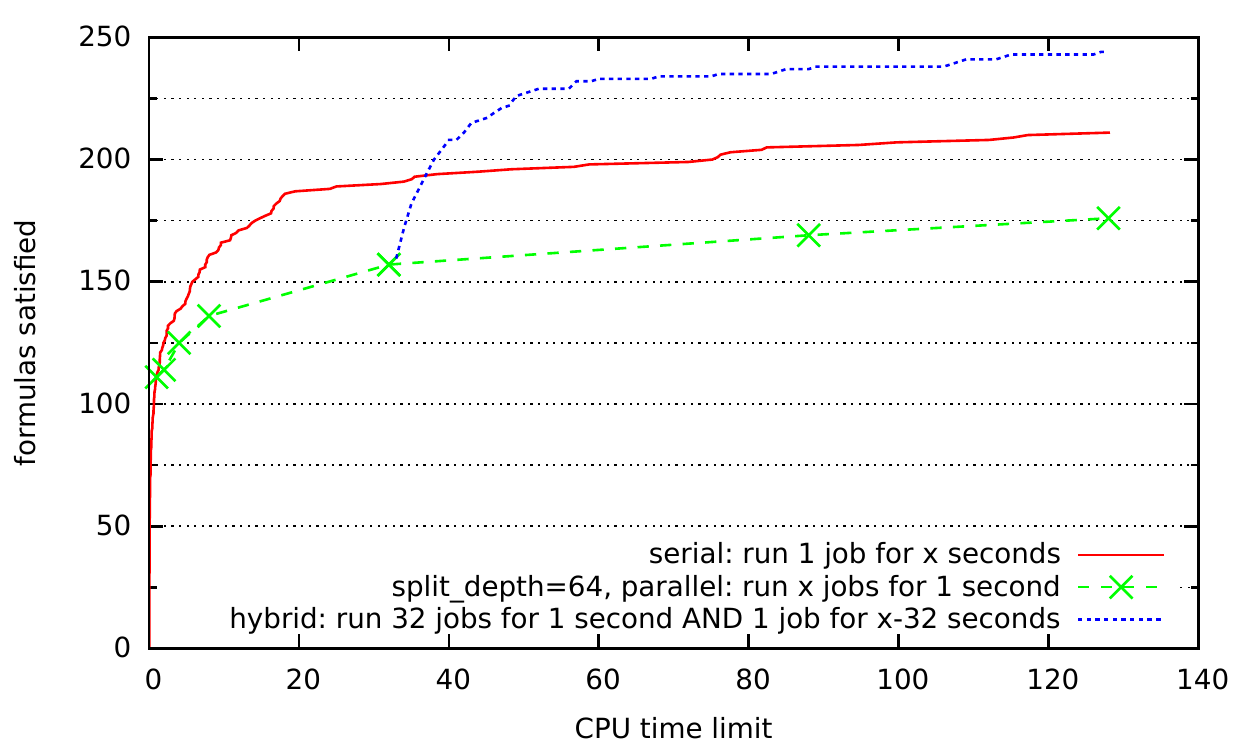}}
		\caption{\label{fig:satgraph}Number of formulas in the benchmark set shown to be satisfiable within $x$ CPU-seconds (e.g. number of CPUs allocated times time limit in seconds).}
	\end{center}
\end{figure}

Our coarse grained parallelism does not assist in processing a single
branch faster. Rather, it allows {\tt leviathan} to process multiple branches
at the same time. Broadly, there are two possibilities: either one
of alternate branches finds a solution faster, or the tableau heuristics
picked the best branch anyway. For this reason, with satisfiable formulas
we do not expect the performance to scale smoothly. Adding additional
CPUs may merely add a little overhead, or it may make an otherwise
infeasible problem suddenly feasible, as in Table \ref{tab:Performance}.

We consider the average performance over the benchmark sets in
Table \ref{tab:Speedup-factor}. We see that in each of the considered
satisfiable cases, the median speedup factor is close to one; suggesting
that parallel algorithm 
is not typically faster than the serial
algorithm. 
On the other hand, the mean speedup is impressive.

The difference between the median and the mean is explained due the
to fact that there is not limit to how much faster the parallel algorithm
can be. If the serial algorithm finds a model, then at least one of
the parallel jobs can find the model just as easily.
\iflong{
 Thus, the lower
median is purely due to the cloud vCPUs being slower than the
local i7, and a small amount of overhead in delegating the tasks across
a cluster. Across all the formulas, this gave a minimum speedup factor
of 0.50.}
On the other hand if the serial algorithm guesses wrongly
as to which path leads to a model, it can perform very poorly. If
one of the parallel jobs picks a better path, it could solve
the problem almost instantly.

Given that the speedup on satisfiable formulas as we increase the number of CPUs is not smooth, we may instead consider the number of formulas that can be shown to be satisfiable within a single second. In Figure \ref{fig:satgraph} we present the total number of formulas in the benchmark sets considered that can be shown to be satisfiable. 
A {\tt split\_depth} of 64 was identified as being effective for showing satisfiability during the pilot study \cite{parallel-long} and was used in this graph.
Increasing the number of jobs increases the number of formulas that are shown to be satisfiable, but not as quickly as increasing the number of seconds allowed. When limiting the serial algorithm to 15 seconds per formula, it can show 176 of the benchmark formulas are satisfiable, the same as the parallel algorithm can when limited to 128 jobs and 1 second. We see that the parallel algorithm is faster but is not as an efficient use of CPU time. On the other hand, the problems it finds hard are different to those the serial algorithm finds hard, so dividing a fixed CPU time budget between the serial and parallel algorithms can be more efficient than the original serial algorithm alone.

\subsection{\label{sec:unsat}Unsatisfiable formulas}

For unsatisfiable formulas the whole tableau must be constructed. Unlike satisfiable formulas, the amount of work will not be effected by the parallelisation changing the search order. For unsatisfiable formulas we expect a smoother increase
in performance as we increase the number of jobs. We see in Table
\ref{tab:Performance} that for unsatisfiable formulas
more jobs consistently provides more performance. We see that out
of this benchmark set the more challenging formulas got close to linear
improvement in performance when increasing from 1 job to 88 jobs.
These formulas were primarily N5x formulas from the {\tt trp} set of randomly generated formulas.

Although tableaux are generally less useful for showing unsatisfiability,
the \texttt{pltl graph}\footnote{Available, (1) of \url{http://users.cecs.anu.edu.au/~rpg/PLTLProvers/}}
tableau based on the Schwendimann \cite{DBLP:conf/tableaux/Schwendimann98}
technique is effective for showing unsatisfiability of N5x formulas.
The \texttt{pltl} tableau took 21.7 seconds to show unsatisfiability of N5x
formulas compared with 14.4 for the fastest technique, putting \texttt{pltl}
in second place. Thus parallelising tableau for showing unsatisfiability
of N5x formulas may lead to practical performance improvements in
this area. Due to the interdependence of branches, parallelising \texttt{pltl}
will not be as easy as parallelising {\tt leviathan}, however.

\section{\label{sec:Analysis}Analysis}
It this section we will discuss why some formulas are hard to parallelise, in context of the shape of the tableaux, and give an intuition of the shape of the tableau to better understand {\tt split\_depth}'s effect.

\subsection{Depth and Width}

\begin{figure}
\includegraphics[scale=0.63]{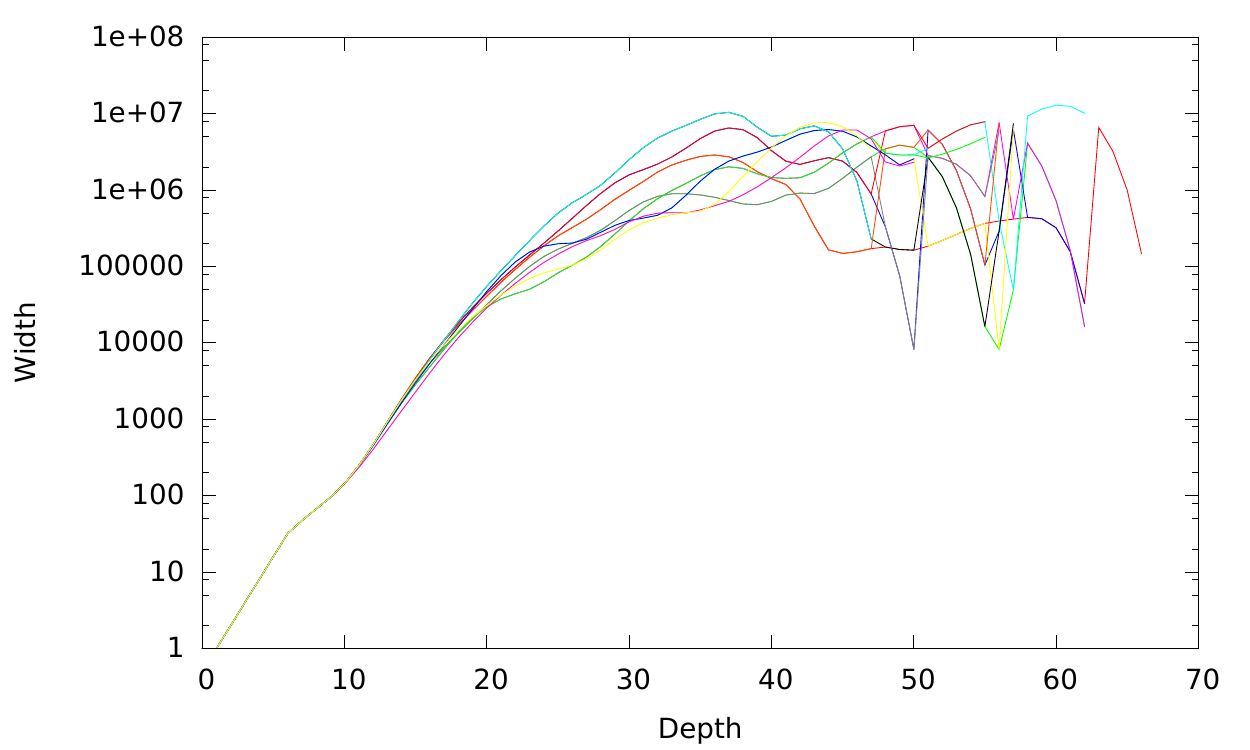}\includegraphics[scale=0.63]{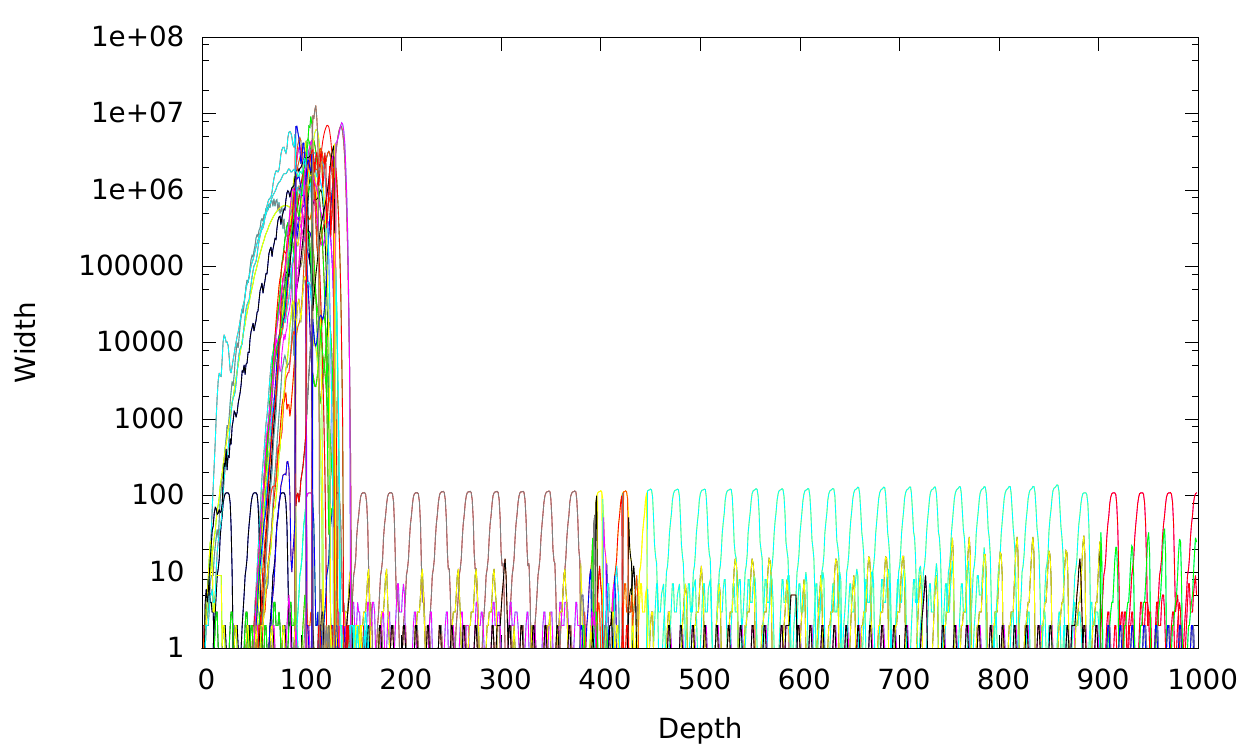}

\caption{\label{fig:Depth-vs-Width}Depth vs Width of tableaux for unsatisfiable
(left: U3) vs satisfiable (right: S3) formulas. Based on tableau constructed
by serial version of {\tt leviathan}. Each line describes the width of the tableau for a different input formula. The width is computed at the time the algorithm terminates, if the formula is satisfiable this may be before the full tableau is constructed. }
\end{figure}

 We consider the problem of finding the ideal \texttt{split\_depth}, and explain
why we chose a default of 18. 

To begin with, let us consider the width of the tableau. This is presented
in Figure \ref{fig:Depth-vs-Width}. In the case of unsatisfiable
tableaux, the width puts an upper bound on the amount of parallelization
that can occur. Fortunately, for the challenging problems in the set
U3, that is quite large. We see in Figure \ref{fig:Depth-vs-Width}
that the depth of the tableau peaks around 1\textendash 10 million
for each of the formulas. If we hope to utilize $n$ CPUs, we must
pick a \texttt{split\_depth} that provides at least that width. We may wish
to have a greater depth, in the hope that if many vertices are assigned
to a single job, that the harder and easier vertices will average out
to provide a roughly equal amount of work in each vertex. 

Increasing the \texttt{split\_depth} also increases overhead. In our algorithm,
all cores have to compute all vertices shallower than \texttt{split\_depth}. In
Table \ref{tab:split-depth-overhead} we investigate how this overhead
grows with increases in \texttt{split\_depth}. The overhead of 1.077 seconds for U3\_007 at a \texttt{split\_depth} 25 is trivial
for a small number of vertices, since the total time is about 80 seconds. However, we are interested in the capability of tableaux
to be massively parallelised, and the overhead becomes significant over
100 jobs, and guarantees that no matter how many jobs are used, the run time
could never be brought down to 1 second with a {\tt split\_depth} of 25.
\begin{table}
\centering{}%
\begin{tabular}{|l|r@{.}l|r@{.}l|r@{.}l|r@{.}l|r@{.}l|r@{.}l|r@{.}l|r@{.}l|r@{.}l|r@{.}l|r@{.}l|}
	\cline{2-23}
	\multicolumn{1}{}{}&\multicolumn{22}{|c|}{\tt split\_depth}\\
	\hline
	Name&\multicolumn{2}{|c|}{15}&\multicolumn{2}{|c|}{16}&\multicolumn{2}{|c|}{17}&\multicolumn{2}{|c|}{18}&\multicolumn{2}{|c|}{19}&\multicolumn{2}{|c|}{20}&\multicolumn{2}{|c|}{21}&\multicolumn{2}{|c|}{22}&\multicolumn{2}{|c|}{23}&\multicolumn{2}{|c|}{24}&\multicolumn{2}{|c|}{25}\\
	\hline
	U1\_001 & 0&001 & 0&002 & 0&002 & 0&002 & 0&003 & 0&003 & 0&004 & 0&004 & 0&005 & 0&007 & 0&008\\
	U2\_001 & 0&006 & 0&010 & 0&017 & 0&030 & 0&050 & 0&080 & 0&127 & 0&184 & 0&251 & 0&338 & 0&425\\
	U3\_001 & 0&007 & 0&012 & 0&019 & 0&033 & 0&054 & 0&084 & 0&134 & 0&205 & 0&303 & 0&465 & 0&657\\
	U3\_002 & 0&006 & 0&011 & 0&017 & 0&028 & 0&043 & 0&066 & 0&106 & 0&159 & 0&236 & 0&353 & 0&489\\
	U3\_003 & 0&007 & 0&011 & 0&019 & 0&033 & 0&055 & 0&089 & 0&144 & 0&218 & 0&324 & 0&489 & 0&703\\
	U3\_004 & 0&007 & 0&010 & 0&016 & 0&027 & 0&043 & 0&067 & 0&102 & 0&139 & 0&177 & 0&228 & 0&288\\
	U3\_005 & 0&006 & 0&010 & 0&017 & 0&029 & 0&049 & 0&082 & 0&143 & 0&231 & 0&362 & 0&529 & 0&698\\
	U3\_006 & 0&006 & 0&008 & 0&013 & 0&023 & 0&037 & 0&058 & 0&094 & 0&140 & 0&204 & 0&301 & 0&417\\
	U3\_007 & 0&006 & 0&011 & 0&018 & 0&033 & 0&056 & 0&096 & 0&166 & 0&283 & 0&427 & 0&697 & 1&077\\
	U3\_008 & 0&006 & 0&011 & 0&018 & 0&030 & 0&046 & 0&071 & 0&106 & 0&150 & 0&205 & 0&280 & 0&353\\

	\hline
\end{tabular}
\caption{\label{tab:split-depth-overhead}Total overhead of \texttt{split\_depth} over
all U0\textendash U3 (unsatisfiable) formulas, measured in seconds
taken by Job 0 (no vertices assigned to job)}
\end{table}
We will estimate how long the parallel tasks will take by recording
how long it took the serial task to process each vertex and its children.
We can determine which vertex would be assigned to which job, add up
the amount of time taken to process all vertices assigned to that job.
This allows us to estimate the length of each job, the amount of time taken to show unsatisfiability is roughly the length
of the longest job.
\begin{table}
{%
\begin{center}
\begin{tabular}{|l|r@{.}l|r@{.}l|r@{.}l|r@{.}l|r@{.}l|r@{.}l|r@{.}l|r@{.}l|}
	\cline{2-17} 
	\multicolumn{1}{c|}{} & \multicolumn{8}{c|}{Raw Estimate} & \multicolumn{8}{c|}{Actual time}\\
	\hline
	Name&\multicolumn{2}{|c|}{job 1}&\multicolumn{2}{|c|}{job 2}&\multicolumn{2}{|c|}{job 3}&\multicolumn{2}{|c|}{job 4}&\multicolumn{2}{|c|}{job 1}&\multicolumn{2}{|c|}{job 2}&\multicolumn{2}{|c|}{job 3}&\multicolumn{2}{|c|}{ job4}\\
	\hline
	U3\_0001 & 6&27 & 6&62 & 6&15 & 6&46 & 6&32 & 6&33 & 6&21 & 6&19\\
	U3\_0002 & 9&14 & 9&14 & 6&78 & 6&65 & 6&65 & 6&66 & 6&87 & 6&73\\
	U3\_0003 & 11&92 & 11&87 & 11&69 & 12&32 & 12&18 & 11&84 & 11&97 & 12&12\\
	U3\_0004 & 12&94 & 13&01 & 13&20 & 12&43 & 12&80 & 13&20 & 13&08 & 12&63\\
	U3\_0005 & 14&19 & 13&93 & 13&98 & 13&81 & 14&23 & 14&25 & 14&29 & 14&10\\
	U3\_0006 & 17&40 & 19&03 & 18&30 & 18&26 & 18&08 & 19&45 & 19&01 & 18&61\\
	U3\_0007 & 20&66 & 20&06 & 20&02 & 21&04 & 21&40 & 20&65 & 20&76 & 21&65\\
	U3\_0008 & 30&27 & 29&17 & 30&22 & 26&44 & 30&89 & 29&86 & 30&68 & 27&02\\
	\hline
\end{tabular}
\end{center}
}

\caption{\label{tab:estimate4CPU-1}Time taken by to process each formula,
divided into four jobs at depth 17, using the ``{\tt to\_job}'' algorithm
to assign vertices to jobs. Left is the raw estimate in seconds based on the number
of time taken processing vertices in the single threaded case. 
 }
\end{table}

We now consider what \texttt{split\_depth} is a reasonable default in Table
\ref{tab:Estimated-performance-at-d}. We see that, depending on the number of jobs,  18--20 is reasonable for the unsatisfiable formulas considered.

\subsection{When parallelisation does not help}


The difficulty in parallelising the other rows of Table \ref{tab:Performance} stems from the narrow
width of the tableau. Many formulas only had a width of one, and in
other cases the width of the tableau continually returned to one,
indicating that only one branch was particularly difficult (see Figure
\ref{fig:Depth-vs-Width} for a visualisation). 

Brute force parallelisation does not always substitute for finding
a more appropriate algorithm. For example, \texttt{pltl graph} takes just under
 4 seconds to show that trp/N5x/12/pltl-5-0-12-3-0-200003 is unsatisfiable.
With a split depth of 26 overhead was 3.9--4.1s, so clearly we cannot 
beat \texttt{pltl graph} on the problem using a split depth greater than 26.
We have 224984 vertices at depth 26, so we could in principle have up
to 224984 jobs running at the same time. However, Task 2 out of 224984 did not complete within minutes, let alone seconds. 

\section{\label{sec:compare}Parallel Tableaux as Contributor to Portfolio Reasoners}

\begin{table}

		\begin{tabular}{|l|r@{.}l|r@{.}l|r@{.}l|r@{.}l|r@{.}l|r@{.}l|r@{.}l|r@{.}l|r@{.}l|r@{.}l|r@{.}l|}
			\cline{2-23}
			\multicolumn{1}{}{}&\multicolumn{22}{|c|}{\tt split\_depth}\\
			
			\hline
			jobs&\multicolumn{2}{|c|}{14}&\multicolumn{2}{|c|}{15}&\multicolumn{2}{|c|}{16}&\multicolumn{2}{|c|}{17}&\multicolumn{2}{|c|}{18}&\multicolumn{2}{|c|}{19}&\multicolumn{2}{|c|}{20}&\multicolumn{2}{|c|}{21}&\multicolumn{2}{|c|}{22}&\multicolumn{2}{|c|}{23}&\multicolumn{2}{|c|}{24}\\
			\hline
			1 & 527&0 & 527&0 & 527&1 & 527&1 & 527&3 & 527&5 & 527&9 & 528&6 & 529&5 & 527&5 & 528&0\\
			2 & 265&5 & 268&8 & 270&3 & 270&4 & 267&2 & 270&8 & 268&9 & 267&7 & 268&5 & 267&6 & 268&4\\
			4 & 142&2 & 137&4 & 140&0 & 139&4 & 137&7 & 138&3 & 136&6 & 137&5 & 138&2 & 136&6 & 138&8\\
			8 &  74&6 &  73&2 &  73&4 &  71&4 &  73&1 &  72&9 &  71&0 &  71&9 &  74&0 &  72&2 &  73&3\\
			16 &  43&1 &  41&4 &  39&8 &  39&0 &  37&9 &  37&9 &  38&6 &  39&4 &  40&4 &  39&3 &  40&9\\
			32 &  25&3 &  22&7 &  21&5 &  21&6 &  21&0 &  21&6 &  21&9 &  21&9 &  23&4 &  22&7 &  24&3\\
			64 &  14&9 &  14&2 &  12&7 &  12&6 &  13&3 &  12&7 &  12&5 &  13&4 &  14&8 &  14&3 &  16&1\\
			88 &  12&2 &  10&5 &  10&8 &  10&8 &  11&2 &   9&8 &  10&2 &  11&4 &  12&3 &  12&1 &  13&4\\
			128 &   9&5 &   8&7 &   8&4 &   8&3 &   7&6 &   8&1 &   8&6 &   9&4 &  10&5 &  10&1 &  11&5\\
			\hline
		\end{tabular}
		\
		\caption{\label{tab:Estimated-performance-at-d} Estimated time to solve all the unsatisfiable formulas in sets U0--U3 performance given various number of parallel jobs and choices of {\tt split\_depth} }		
	\end{table}
	
	In this section we consider what part a parallel tableau could play
as part of a combined tool
which uses several more basic 
LTL satisfiability techniques 
as alternatives in parallel on a given input.
For example, the portfolio tool
POLSAT \cite{DBLP:journals/corr/LiP0YVH13},
makes use of a range
of the best tableaux, resolution,
automata and symbolic tools,
all set off in parallel on the input formula.

Also published in 
\cite{DBLP:journals/corr/LiP0YVH13}
is a comprehensive
list of benchmarking
a variety of best performing reasoners
across
 the families of Schuppan benchmarks
 (\url{http://www.schuppan.de/viktor/atva11/}). 
 From this we can see
 that the tableau tools
 are often the
 fastest type of tools across satisfiable formulas
 but rarely on unsatisfiable formulas.
 
 From the results in Sections \ref{sec:sat} and \ref{sec:unsat}
 above
 we also know that
 parallelising suitable
 tableaux (such as the Prune-rule-based Leviathan approach)
 can lead to 
 (1) easily gained impressive speed-ups on some satisfiable formulas;
 (2) not much of a speed-up on many satisfiable formulas; and
 (3) solid speed-ups
 on many unsatisfiable formulas.
 Together these observations
 suggest that
 it is worth
 incorporating a parallel tableau approach
 in these portfolio approaches.

We should also mention the very recent new
SAT-based explicit temporal reasoner presented in
\cite{Li2015}
which uses the technique of
``semantic splitting'' \cite{MaZ2009}
like 
state-of-the-art propositional SAT solvers, 
instead of the syntactic splitting
seen in tableaux approaches.
This technique is showing impressively fast results via the implementation
Aalta v2.0.
The timings presented in the appendix
of \cite{Li2015}
show that this is much faster on most of the benchmarks than
LS4 \cite{SuW2012},
TRP++ \cite{hustadt2003trp++},
NuXmv-BMCINC \cite{HJL05},
and
the tableaux/automata reasoner 
Aalta v1.2 from \cite{DBLP:journals/corr/LiP0VH14}.
 Aalta 2.0 does seem to perform significantly faster than the other tools
 almost across the benchmarks
 so it will be 
 interesting to see in future work whether the speed-ups that
 we are seeing in parallel tableaux techniques 
 can challenge this in any way.

 \subsection{Effectiveness}
We are most interested in comparing the performance of the parallel tableau to other tableaux. The tableaux {\tt pltl graph} and {\tt pltl tree} are available from  \url{http://users.cecs.anu.edu.au/~rpg/software.html}. These are particularly interesting because {\tt pltl graph} is widely considered the most promising LTL tableau, and {\tt pltl tree} is a tree tableau similar to {\tt leviathan}. 
 
 To evaluate the intuition that a parallel {\tt leviathan} would be a useful contribution to a portfolio reasoner we generated a large number of new pseudo random formulas. Although the {\tt trp} set contains randomly generated formulas, the set of {\tt trp} formulas that are not already trival to solve with existing serial tableau techniques is too small to form a meaningful benchmark set.
  
 We adapt our existing random generator for Full Computation Tree Logic (CTL{*}) \cite{MDR:rewlong} formulas to LTL, resulting in the following
 recursive procedure: we generate a formula of length $n$ recursively
 as follows. For $n=1$ we pick an atomic proposition at random from
 $\left\{ p,q\right\} $; for $n=2$ we pick at random a unary operator
 from $\{\neg,\X,\F,\G\}$ and an atom from $\left\{ p,q\right\} $.
 For $n>2$ we first choose to start with a binary operator or unary
 operator with equal probability.
  Out of $10^6$ formulas of length 50, we eliminated all formulas that could be
 reasoned with by {\tt pltl graph} or {\tt pltl tree} within a second. This left 2779 formulas. Eliminating the formulas that could not be solved by serial {\tt leviathan} left 314 unsolved formulas. Using 8 jobs with a {\tt split\_depth} of 18 or 20 reduced the number of unsolved formulas to 300 or 299 respectively, a slight improvement over the serial case.

\iflong
{  
This left 4693 formulas. We then eliminated all formulas that could
 be solved by {\tt leviathan} within 0.1 seconds, as there was little point
 in parallelising these. This left 581 formulas. Of this 55 were solved
 with 16 CPUs in under 0.1s a {\tt split\_depth} of 16; 43 at 18; 41 at 19;
 33 at 24; 21 at 40; 16 at 48; 15 at 56; 15 at 64; 14 at 72; 11 at
 80. Slightly more could be solved in one second (67@16 56@18 54@19
 47@24 31@40 27@48 27@56 24@64 24@72 23@80). 104 formulas could be
 solved within a second using at least one of the considered split
 depths, and of those 91 could not have been solver by the original
 serial {\tt leviathan} within a second. We see that parallelisation can
 easily decrease the number of problems not solved by any of these
 tableau solvers by 10-20\%. 
}

PolSAT is presently available from \url{http://lab301.cn/home/pages/lijianwen/}. We were not able to get the bundled version of {\tt altaa} to compile, so we replaced it with version 2.  It was difficult to find formulas of length 50 that could not already be solved within 1 second by PolSAT, so we increased the formula length to 400. We found 23460 formulas that could not be solved by PolSAT in 1 second. Out of these, {\tt leviathan} could solve 100 in one second (0.5\%).
We took 2000 formulas that had not been solved in one second by either the serial leviathan tableau or PolSAT. We then put those 2000 formulas into the parallel algorithm running on the cloud with 16 vCPUs (8 physical CPUs), using a {\tt split\_depth} of 20. The parallel algorithm was able to show that 6 (0.3\%) of these formulas were satisfiable, and was not able to show that any were unsatisfiable. The hardest of these 6 formulas took PolSAT 28 seconds. 

In summary, although parallelising a tableau algorithm can provide dramatic speedups, novel algorithms have more to offer than parallelisation of existing algorithms.

\section{Conclusions and Future Work}
We see that parallelising tableaux can indeed be easy. Though our parallelisation technique is simple, its greatest weakness is simply that a highly parallel tableau is no substitute for a tableau optimised to the class of LTL formulas of interest. Thus, the shortage of prior parallel tableau may be explained by the vast potential for performance improvements in serial algorithms.
We have demonstrated a significant
speedups over the serial version. These speedups are not guaranteed
for satisfiable formulas, but are frequently much better than linear.
For unsatisfiable formulas, the speedups are reliable, but sublinear. When it is reasonable to use a large number of CPUs, this provides a significant incremental improvement in the amount of formulas considered that can be reasoned about using tableau.

We have also shown that this simple approach is far from
a panacea.  
In general, {\tt leviathan} is highly effective at showing that  a particular class of formulas is satisfiable, and parallelisation broadens the class of formulas it can show is satisfiable. Even massive parallelisation would not make {\tt leviathan} beat techniques optimised for showing unsatisfiability at their own game.

It is not clear that more sophisticated parallelisation techniques would help. Before settling on the present algorithm, we tried starting a new process for each branch at {\tt split\_depth} so that we could dynamically assign branches to CPUs as they became idle. The overhead of starting new processes turned out not to be worthwhile.
One could use a POSIX ``{\tt fork()}'' to avoid this overhead, but this would not scale naturally past a single machine.  One could instead serialise the state of the tableau and transmit it to a new machine, but this would be invasive as {\tt leviathan} has various types of internal state. 

Since we have had close to linear growth in performance in unsatisfiable formulas, there is little to gain here by improving parallelisation. By contrast switching to a tableau better at showing unsatisfiability such as {\tt pltl} {\tt graph} gave massive improvements in
 of some formulas. Simply improving the parallelisation of {\tt leviathan} will not bring its performance close to 
  {\tt pltl} {\tt graph} on those formulas.  

This simple parallelisation technique is more practically useful for making {\tt leviathan} even more effective at showing satisfiability than it already is. In the case of the more challenging satisfiability formulas in the S3 set and a {\tt split\_depth} of 20, there were no cases a job terminating before the formula was shown to be satisfiable, so there were no idle CPUs to move work to. Thus, it is hard to justify the complexity of more advanced parallelisation techniques. This complexity would be better spent on more advanced heuristics and parallelising more reasoning techniques. 
We have seen that adding a parallel version of {\tt leviathan} would only improve PolSAT slightly. We would expect a greater increase from  parallelising the more optimised {\tt pltl} {\tt graph} tableau, a more challenging task.


The unsatisfiable formulas studied had a fairly similar exponential growth in up to depth 20. This provided enough width for a high degree of parallelism over these formulas at the fixed \texttt{split\_depth} of 18. By contrast satisfiable formulas were less predictable as to where the ideal \texttt{split\_depth} was, frequently 64 was better than 18 for satisfiable formulas. There would be some benefit to automatically detecting the ideal {\tt split\_depth}, however for a number of formulas no fixed depth would suit and a different method of parallelisation would need to be used.

{\bf Acknowledgements. } This research was funded partially by the Australian Government through the Australian Research Council. (project DP140103365)

\bibliographystyle{eptcs}
\bibliography{csbibtex,mark,dblp}

\end{document}